\documentclass[12pt,preprint]{aastex}
\pdfoutput=1
\shorttitle{MODELING OF UPFLOW IN DIMMING REGION}
\shortauthors{Imada et al.}

\begin{document}

\title{ 1D MODELING FOR TEMPERATURE-DEPENDENT UPFLOW IN THE DIMMING REGION OBSERVED BY HINODE/EIS}

\author{S. \textsc{Imada},\altaffilmark{1} 
H. \textsc{Hara}, \altaffilmark{2}
T. \textsc{Watanabe}, \altaffilmark{2}
I. \textsc{Murakami},\altaffilmark{3}
L.K. \textsc{Harra}, \altaffilmark{4}
T.\textsc{Shimizu}, \altaffilmark{1}
E. G. \textsc{Zweibel}\altaffilmark{5} 
}
\altaffiltext{1}{ Institute of Space and Astronautical Science, Japan Aerospace Exploration Agency, 3--1--1 Yoshinodai, Chuo-ku, Sagamihara-shi, Kanagawa 252--5210, Japan}
\altaffiltext{2}{ National Astronomical Observatory of Japan,
  2--21--1 Osawa, Mitaka-shi, Tokyo 181--8588, Japan}
\altaffiltext{3}{ National Institute for Fusion Science,
  322--6 Oroshi-cho, Toki, Gifu 509--5292, Japan}
\altaffiltext{4}{UCL-Mullard Space Science Laboratory, Holmbury St Mary, Dorking, Surrey, RH5 6NT, UK}
\altaffiltext{5}{University of Wisconsin-Madison, 475 N Charter Street, Madison, WI 53706, USA}

\begin{abstract}
We have previously found a temperature-dependent upflow in the dimming region following a coronal mass ejection (CME) observed by the  {\it Hinode} EUV Imaging Spectrometer (EIS). 
In this paper, we reanalyzed the observations along with previous work on this event, and provided boundary conditions for modeling. 
We found that the intensity in the dimming region dramatically drops within 30 minutes from the flare onset, and the dimming region reaches the equilibrium stage after $\sim$1 hour later.
The temperature-dependent upflows were observed during the equilibrium stage by EIS. 
The cross sectional area of the fluxtube in the dimming region does not appear to expand significantly. 
From the observational constraints, we reconstructed the temperature-dependent upflow by using a new method which considers the mass and momentum conservation law, and demonstrated the height variation of plasma conditions in the dimming region. 
We found that a super radial expansion of the cross sectional area is required to satisfy the mass conservation and momentum equations.
There is a steep temperature and velocity gradient of around 7 Mm from the solar surface.
This result may suggest that the strong heating occurred above 7 Mm from the solar surface in the dimming region.
We also showed that the ionization equilibrium assumption in the dimming region is violated especially in the higher temperature range.
\end{abstract}

\keywords{MHD---plasmas---Sun: corona---Sun: flare---Sun: UV Radiation}

\section{INTRODUCTION}
A solar flare is often explained as a rapid energy conversion phenomenon from magnetic field energy to plasma energy. 
The released energy by a flare is so huge, and the total amount of energy often reaches to $10^{32}$ erg within an hour. 
Over the past few decades, considerable effort has been devoted toward understanding the physical mechanism of solar flare, and various models have been proposed (e.g., \cite{car, stu,hir,kop}). 
Nowadays, it is widely accepted that magnetic reconnection is regarded as the fundamental energy conversion mechanism of eruptive flares. 
So far, various features expected from the model have been discussed and confirmed by the modern observations \citep[e.g.,][]{tsu,ohy,yok,ter}. 

Another important aspect of the solar flare is its impact on surroundings.
Most powerful flares are associated with a Coronal Mass Ejection (CME)  \citep[e.g.,][]{kah,yas}. 
Because a CME releases a huge amount of plasma and magnetic field into the interplanetary space in a short duration of time, the equilibrium in the solar corona breaks down. 
Therefore, large-scale magnetic reconfiguration in the solar corona occurs. 
The opposite scenario also can work; a large-scale magnetic reconfiguration can cause a CME. 
Although the causal relationship between large-scale magnetic reconfigurations and CMEs is still not clear, a solar flare have a large impact on its surrounding through CME or large-scale magnetic reconfiguration.

Coronal dimming in the EUV and/or soft-Xray range \citep[e.g.,][]{rus, ste, tho}, which is due to a loss of coronal plasma, is often observed as one of the main on-disk signatures of a CME. The dimming can persist longer than a day in the some extreme cases, and its mass loss contribution can reach more than 50\% of the total mass of the CME \citep[e.g.,][]{ste}. 
There are two scenarios that can explain the origin of dimming regions. 
One is the eruption of a flux rope, and the other is interchange reconnection between closed and open magnetic field lines.
\cite{ste} found a pair of dimming regions which are located at either end of a preflare sigmoidal (S-shaped) structure. They concluded that their findings are consistent with the source of the CME being a flux rope that erupted, leaving behind the dimming regions.
\cite{att} discussed the magnetic field topology changing associated with a CME. In this scenario, the magnetic loops of the CME expand and eventually push against the oppositely oriented open magnetic field of the polar coronal hole. This triggers successive magnetic reconnection, and some part of the magnetic loop topology changes from closed to open.
Although it is not clear which is the dominant process, in either case the dimming region is likely to be a footpoint of open field line. The changing magnetic configuration from closed to open within a short period causes the evacuation of coronal material towards interplanetary space through the rarefaction wave, and plasma cannot refill the open magnetic field lines for a long time.

Quantitative measurements of dimming regions are essential to understand its contribution to CME formation or development.
The EUV imaging spectrometer (EIS) on board {\it Hinode} \citep{kos} provides EUV spectra with higher spectral and spatial resolution than ever before \citep[][]{cul}.
The first measurement of dimming regions by EIS was the flares and CMEs of 13 and 14 December 2006 in NOAA 10930. 
\cite{har} studied the outflows in the dimming region of the 2006 December 14 events, and they found for the first time that the strong outflow exists at the loop footpoints in the dimming region which is located far away from the flare site. 
\cite{ima} analyzed the flare of 13 December 2006 and found a temperature dependence of the outflows; the hotter plasma shows the faster upflow. 
They claimed that the temperature-dependent upflows are the consequence of coupling between heating and bulk acceleration along the line-of-sight (height direction). \cite{jin} have analyzed both of these December 2006 events and found that the velocity also correlates with the photospheric magnetic field. \cite{har2} discussed the magnetic flux from the source of the CME by using the Doppler  velocity difference and the photospheric magnetic field measurement. They found that their estimation was consistent with the magnetic flux in the magnetic cloud observed at the interplanetary medium (1 AU) by in situ measurement.

In this paper, we explore further the temperature-dependent upflow in the dimming region observed by {\it Hinode}/EIS.
The aim of our paper is to reconstruct the temperature-dependent upflow by using a new method which uses the mass and momentum conservation law to determine the height variation of the plasma conditions in the dimming region. This paper is organized as follows. In the next section, we discuss the observational properties in dimming region of the  2006 December 13 event. Section 3 is devoted to the modeling of temperature-dependent upflow.  We discuss the time-dependent ionization effect in Section 4.  Summary and discussion are given in \S 5.

\section{Observation of dimming region }

In this section, we discuss the observational properties in the dimming region on the 2006 December 13 event.
We have reanalyzed the event to obtain quantitative information to reconstruct the dimming region which is discussed in the next section.
We also discuss the previous observational results of the same event.

\subsection{Instrumentation}
The EIS instrument  onboard {\it Hinode} is a high spectral resolution spectrometer aimed at studying dynamic phenomena in the corona with high spatial resolution and sensitivity. 
While making a large raster scan with 1 arcsec slit from 01:12 UT to 05:42 UT on 2006 December 13, EIS observed an X3.2 flare.
The flare occurred at 02:14 UT accompanied by the halo CME.
The flare itself has been studied in detail \citep[e.g.,][]{kub, ima2, asa, min}.
EIS data from the raster were processed using the EIS team provided software (EIS\_PREP), which corrects for the flat field, dark current, cosmic rays, and hot pixels. 
Each spectrum was fitted with a Gaussian profile, and the line width and Doppler velocity were determined.
The slit tilt was corrected by eis\_tilt\_correction. For thermal reasons, there is an orbital variation of the line position causing an artificial Doppler shift of +/- 20 km s$^{-1}$ which follows a sinusoidal behavior. This orbital variation of the line position was  corrected using the method described by \cite{ima}, and the instrumental effects are reduced to below +/- 5 km s$^{-1}$, as a result.

For studying the time evolution of dimming region, we made use of the Extreme ultraviolet Imaging Telescope (EIT) on SOHO \citep[][]{del}. We used the EIT images in the 195  \AA~ filter. The EIT data have a cadence of 12 min and a pixel size of 2.6 arcsec pixels. The exposure times are $\sim$12.6 sec during the event. We processed EIT data by using eit\_prep for the calibration.

\subsection{SOHO/EIT observation}
An EIT base difference image using the 195 \AA~ filter is shown in Figure 1. 
A pre-flare image at 02:00 UT was subtracted from the post-flare image at 03:36 UT. 
The flare site is located around (350, -150), and there is a deep dark dimming region in the far east from the flare site. 
We highlighted the dimming region  in which the intensity drops by more than 60\% of its original intensity with a red contour. 
The time evolution of the dimming region can be clearly seen in Figure 2. 
Most loops or diffuse structures inside the red contour were blown off within the first 30 minutes, and the bright point-like structures, which may be related to foot points of coronal loops, loose their intensity gradually. 
Figure 3 shows the time evolution of the average intensity in dimming region (inside the red contour in Figure 1). 
We can clearly see that the intensity is rapidly decreasing and reaching the quasi-equilibrium stage. 
Similar features were observed in the event on the next day.  \cite[e.g.,][]{har}.
Long term evolution of this dimming region is discussed by \cite{att2}.
They found that concentrated downflows develop during the recovery phase of the dimmings and are also correlated with the same magnetic elements that were related to outflows during the quasi-equilibrium stage.

\subsection{Hinode/EIS observation}
EIS observed the dimming region from 04:00 to 05:30 UT, and obtained the spectral information in the several emission lines.
During this period, the dimming region already reached the quasi-equilibrium stage (see Figure 3).
In Figure 4, we show the intensity, velocity, and line width map of \ion{He}{2} (256.32~\AA; $\log T_{max}=4.70$), \ion{Fe}{12} (195.12~\AA; $\log T_{max}=6.11$), and \ion{Fe}{15}  (284.16~\AA; $\log T_{max}=6.30$) obtained by EIS. 
The color scales in the velocity map range from -150 to 150 km s$^{-1}$.
The color scales in the line width (Full Width at Half Maximum) map are normalized by the median of line width in whole field of view.
The median values of line width in \ion{He}{2}, \ion{Fe}{12}, and \ion{Fe}{15} are 0.102, 0.073, and 0.085 \AA, respectively. 
Note that EIS scans from west (right) to east (left). 
The time stamps of slit scanning are also shown in the top of Figure 4. 
In the intensity map, the bright flare site can be observed in western part, and we can clearly see the dimming region in the far east from the flare site, which is framed by white dashed lines.
The temperature-dependent upflow in the dimming region, which was discussed in \cite{ima} and \cite{jin}, was also clearly seen in the Doppler velocity map in Figure 4. 
The strong outflows ($\sim$ 100 km sec$^{-1}$) in higher temperature are located in the boundary of the dimming region. 
The line width in the dimming region also shows the temperature dependence.
This is because the inhomogeneities of flows in the dimming region produce the spatial variations of Doppler velocities, which result in an apparent broadening of the integrated profile.
Therefore, it is reasonable that the line widths in the dimming region also have temperature dependence. 
The relationship between flows and line broadening are discussed in \cite{ima2} and \cite{dol}.

To be confident of our results we need to understand any possible contamination from line blends. 
Generally, \ion{Fe}{12} and \ion{Fe}{15} lines are very strong in the corona, and we do not need to take into consideration line blending.
On the other hand, in the \ion{He}{2} line, there are strong line blends with the  \ion{Si}{10} which has a formation temperature of $\sim 10^{6.2}$ K \citep[][]{you}.
However, the blend is not significant in the dimming region, because the line intensity from the coronal temperature plasma is very weak  (Figure 4).
Hence, we can neglect the line blend for  \ion{He}{2} in the dimming region.
The line blend in the \ion{He}{2} line is also discussed in \cite{jin}, and they also concluded that  the contribution of the \ion{Si}{10} line is
more significant in the active regions than in the dimming areas.

In the regions where the Doppler velocity is the highest  in \ion{Fe}{15}, we fit the line profiles using a  double gaussian for several ions. 
The line profiles are clearly separated into two components - the fast upflow and the stationary component (not shown here, see \cite{ima}).  
The fast upflow may correspond to the magnetic field whose topology has been changed from close to open by the flare activity.
The other stationary component may correspond to the magnetic field whose topology was not changed by the flare activity.
It is plausible that these two components are mixed within our spatial resolution, because the dimming region locates very far from the flare site.
Figure 5 shows the relationship between the velocity (fast) and temperature in the dimming region. 
The line formation temperatures are defined by the temperature that yields its peak abundance with the CHIANTI code \citep[][]{lan}.
This naturally means that we assumed the ionization equilibrium to derive the line formation temperature.
The line blending effect does not have a significant effect on the velocity estimation in the result \citep[see,][]{ima}.
The result of the relationship between the temperature and upflow velocity is also discussed in \cite{jin}. 
The velocities in our result is much faster than those in \cite{jin}, because we fitted the line profiles by the double gaussian.  
To discuss the relationship between the temperature and velocity quantitatively, we fit it by the following equation;
\begin{equation}
T(v) = a_2v^2+a_1v+a_0,
\end{equation}
where, $T, v,$ and $a_{2-0}$ are the electron temperature (K), upflow velocity (m sec$^{-1}$), and fit coefficients. 
The solid line in Figure 5 shows the fitting result, and the fit coefficients are also shown in Table 1.
The correlation coefficient (r) is 0.95.

The area of the EIS dimming region at different temperatures in the corona are also studied by \cite{jin}.
They estimated the area with an intensity drop larger than 5\% at the eruption phase compared with the pre-eruption phase.
They found that the dimming area weakly depends on its temperature.
From their result, the area increases by a factor of 1.6 between 0.4 and 1.6 MK plasma.
They also discussed the relationship between magnetic filed and upflow, and found that the fast upflows are located at relatively strong magnetic fields.
  
There are two interpretations for the temperature-dependent upflow in the dimming region.
One is that the temperature and velocity depend on the height from the solar surface (line of sight direction).
The other is that the different loops in the dimming region have different temperatures and velocities in sub-resolution size (the multi-strand model).
Both of the scenarios can explain the temperature-dependent upflows in the dimming region. 
The recent observations also support both of the interpretations, \citep[e.g.,][]{jin, rob, war}. 
However, the multi-strand model  generally causes large broadening, distortion, and humps in line profile, \citep[e.g.,][]{uga}.  
Therefore, the line profile may not separate into the two distinct components (fast and stationary), which is observed by EIS in the dimming region. Further, the area of the dimming region also weakly depends on the temperature.
The observations might support the former scenario, because generally the fluxtube expands radially in the solar corona.
For those two reasons, we assume that the temperature and velocity depend on the height from the solar surface in the dimming region, although we do not deny the possibility of the multi-strand model.

\subsection{Summary of  observation}
A summary of the observations is given below.
\begin{itemize}
\item The intensity (\ion{Fe}{12}) in the dimming region dramatically drops within 30 minutes from the flare onset. 
\item The dimming region reaches the equilibrium stage after $\sim$1 hour after the flare onset.
\item EIS observed the temperature-dependent upflows in the dimming region.
\item The dimming region is in the equilibrium stage during the EIS observing period.
\item The temperature dependence of the upflow is characterized by Equation 1.
\item The area of the dimming region weakly depends on the temperature (the expansion factor is $\sim$1.6 between 0.4 and 1.6 MK plasma).
\end{itemize}

\section{Modeling}
\subsection{Method for Modeling}
In this section, we try to reconstruct  the temperature-dependent upflow and determine the height dependence of plasma conditions in the dimming region by using the observational facts discussed in the previous section.
Because the reconstruction will be based on our observations, our model can only apply in the range of $16 < v< 160$ km s$^{-1}$ ($0.4 < T_e< 2$ MK). 
We assumed the temperature-dependent upflow is a quasi-steady, because the flow was observed in the equilibrium stage of the dimming region.
The fluxtube is vertical to the solar surface, and the cross sectional area is expanding with height from the solar surface. 
The fluxtube contains low $\beta$ plasma, and we assumed that cross-field motions are negligible so that a one-dimensional description is adequate.
Neglecting time derivatives in the hydro dynamic equations produces the quasi-steady flow model.
The mass conservation equation, equation of state in fully ionized gas, and momentum conservation equation in one fluid description are given by
\begin{equation}
mn v A = const,
\end{equation}
\begin{equation}
p= 2 n k_B T,
\end{equation}
\begin{equation}
 v \frac{d v}{d s}=-\frac{1}{mn}\frac{dp}{ds} - g_0 \left(\frac{R_s}{s+R_s}\right)^2,
\end{equation}
where n, v, A, p, and T are the electron number density, bulk velocity, cross sectional area, total pressure and temperature, respectively,
m is the mean mass for solar abundances (1.257 times of the hydrogen mass), $k_B$ is the Boltzmann constant, $g_0$ is the solar surface gravity, $R_s$ is the sun radius, and s is the height from the solar surface. We neglected the effect of He or the other minor elements in Equation 3.

In order to describe the flow precisely, we rewrite Equation 4 using Equation 2 and 3 as follows;
\begin{equation}
\left(v - \frac{2k_B T}{mv} \right)\frac{d v}{d s} + \frac{2k_B}{m}\frac{d T}{d s}=- g_0 \left(\frac{R_s}{s+R_s}\right)^2+ \frac{2k_BT}{m}\frac{1}{A}\frac{d A}{d s}. 
\end{equation}
The equation is almost the same equation which was used for the discussion of the solar wind \citep[][]{par2} or the discussion of the stationary siphon flows in  coronal loops \cite[e.g.,][]{cra,orl}. 
In a Parker-type wind, the flow is accelerated by a pressure gradient. The pressure drop is typically due to decreases in both temperature and density. 
In the flow described here, T is increasing with v (see Figure 5). 
Therefore, we anticipate that acceleration requires rapid divergence of the fluxtube area and that the flow is strongly heated. 

Equation 5 has a critical point at $v=[2k_BT/m]^{1/2}$. 
Fortunately, the upflow velocity in the dimming region is less than $[2k_BT/m]^{1/2}$, for example  $[2k_BT/m]^{1/2} = 113$ km sec$^{-1}$ at 1 MK.
We substitute $T(v)$ (Equation 1) into Equation 5, and rewrite it in conservative form as follows;
\begin{eqnarray}
&\frac{1}{2}\left(v^2-v_0^2\right)-\frac{2k_B}{m}\left(\frac{1}{2}a_2\left(v^2-v_0^2\right)+a_1\left(v-v_0\right)+a_0\log\frac{v}{v_0}\right)+\frac{2k_B}{m}\left(T-T_0\right)
\nonumber\\
&=-g_0R_s^2\left(-\frac{1}{s+R_s}+\frac{1}{s_0+R_s}\right)+\frac{2k_B}{m}\int^s_{s_0} T\frac{1}{A}\frac{d A}{d s'} ds',
\end{eqnarray}
where $v_0$, $T_0$, $s_0$, are the velocity, temperature, and height at the base of the coronal loop. 
We have chosen the footpoint velocity and height to be 16 km s$^{-1}$ and 2000 km, respectively.
$T_0$ can be determined by Equation 1. 
We simplify each term in Equation 6 as follows; $S_{L1}-S_{L2}+S_{L3}=-S_{R1}+S_{R2}$.
The left-hand side in Equation 6 ($S_{L1}-S_{L2}+S_{L3}$) can be determined when the velocity is determined, and the first term of right-hand side can be determined when the height from solar surface is determined.
Only the second term of right-hand side ($S_{R2}$) is unknown.
Therefore, we need to make an assumption of cross sectional area ($A$) to evaluate Equation 6.

\subsection{Radial expansion for fluxtube}
The geometry of coronal fluxtubes is believed to fan out from the network boundaries in the photosphere to the corona, forming a canopy like structure \citep[][]{gab}. 
This geometry leads to a varying cross sectional area of every fluxtube along the height from the solar surface.
We assumed the cross sectional area expands radially, and it can be written as follows;
\begin{equation}
A(s)=A_0\left(\frac{s+R_s}{s_0+R_s}\right)^\mu,
\end{equation}
where $\mu$ is the parameter for expansion. The uniform, sub-radial expansion, radial expansion, and super-radial expansion in cross section can be expressed by $\mu=0$, $0<\mu<2$, $\mu=2$, and $2<\mu$, respectively. 
In all these cases, $S_{R2}$ can be written as follows;
\begin{equation}
S_{R2}=\frac{2k_B\mu}{m}\int^s_{s_0} T\frac{1}{s'+R_s}ds'.
\end{equation}
Further, we can evaluate the integration part in $S_{R2}$ with 
\begin{equation}
T_{MAX}\int^s_{s_0} \frac{1}{s'+R_s} ds'>\int^s_{s_0} T(s')\frac{1}{s'+R_s} ds'>T_0\int^s_{s_0} \frac{1}{s'+R_s} ds',
\end{equation}
where $T_{MAX}$ is the maximum temperature in our range. 

The value of each term in the left-hand side in Equation 6 is shown as a function of velocity in Figure 6.
We can clearly see that the total of left-hand side terms always has positive values.
In Figure 7, we show $S_{R1}$ as a function of height by the solid line.
Note that the horizontal axis in Figure 6 and 7 are different.
Let us evaluate $S_{R2}$ in the case of radial expansion ($\mu=2$).
The maximum and minimum value obtained from Equation 9 are shown in Figure 7 by the dashed and dotted lines.
Thus, $S_{R2}$ should be located between the dashed and dotted lines.
This result indicates that the right-hand side ($-S_{R1}+S_{R2}$) is always negative.
Therefore, there is no solution of Equation 6 in the case of radial expansion, because the left-hand side in Equation 6 is always positive. 

\subsubsection{Super-radial expansion: Case I ($\mu = 45$)}
From the slope of $S_{R1}$ and $S_{R2}$ at $s=s_0$, we can estimate the lower limit of $\mu$ from the condition being the positive value in the right-hand side of Equation 6  ($dS_{R2}/ds \geq dS_{R1}/ds$); 
\begin{equation}
\mu \geq \frac{m}{2k_BT_0}\frac{g_0R_s^2}{s_0+R_s} \sim 44.
\end{equation}
Thus we assumed the super-radial expansion ($\mu=45$) for the cross sectional area (Case I).
The dotted-dashed and dotted-dotted-dashed lines in Figure 7 show the maximum and minimum value of $S_{R2}$ ($\mu=45$) evaluating from Equation 9.
This result indicates that the right-hand side is always positive.
Therefore we can solve Equation 6 with $\mu=45$ and obtain the height dependence of plasma conditions in the dimming region.
We numerically perform integration of $S_{R2}$. 

The results for the height dependences of various quantities in the dimming region are shown in Figure 8.
The horizontal axis shows the height from the solar surface, and the vertical axes show the velocity, temperature,  cross sectional area, and density. 
The cross sectional area and density are normalized by those at $s=2000$ Mm (=$s_0$).
The upflow is almost constant until a height of 6 Mm and dramatically accelerated up to 160 km sec$^{-1}$ from 6 to 9 Mm. 
The temperature also shows the same profile as the velocity because of Equation 1.
The cross sectional area expands almost proportional to the height from the solar surface, and it reaches around 1.6 at 9 Mm.
The density is gradually decreasing with the height, and rapidly decreasing from 6 Mm.
The density ratios ($n/n_0$) at 7.5 and 9 Mm are almost 0.14 and 0.07, respectively. 
Note that the temperatures at 7.5 and 9 Mm are around 1.4 and 2 MK, respectively.

\subsection{Empirical Model for fluxtube}
The empirical modeling of the relative size of the fluxtube cross sectional area in the quiet sun was done by \cite{cha}.
They obtained the relationship between temperature and velocity from SOHO/SUMER observations in the quiet sun, and derive  the estimation of the relative size of the fluxtube as a function of temperature.
They also found the derived relative size of the fluxtube can be fitted to a functional form suggested by \cite{dow} and \cite{rab}
\begin{equation}
A(T)=A(T_h)\frac{1}{\Gamma}\left(1+\left(\Gamma^2-1\right)\left(\frac{T}{T_h}\right)^\nu \right)^\frac{1}{2},
\end{equation}
with values of $T_h=10^6$ K, $\Gamma=31$, and $\nu=3.6$.
We applied Equation 11  to Equation 6, and $S_{R2}$ can be written as follows; 
\begin{equation}
S_{R2}=\frac{k_B}{m}\frac{\Gamma^2-1}{T_h^\nu}\int^T_{T_0}T'^{\nu}\left(1+\left(\Gamma^2-1\right)\left(\frac{T'}{T_h} \right)^\nu\right)^{-1}dT'.
\end{equation}
Their discussion is concentrated on the down flows ($\lesssim 10$ km s$^{-1}$) observed in the quiet sun transition region (below 1MK).
Although their situation may be different from the dimming region, their results might include the characteristics of the cross sectional area.

\subsubsection{Strong dependence on temperature: Case II ($\nu=3.6$) }
First, we have applied the parameters suggested by \cite{cha} ($T_h=10^6$ K, $\Gamma=31$, and $\nu=3.6$) to Equation 12, and solved Equation 6.
The results of height dependence in the dimming region are shown in Figure 9.
The figure format is the same as Figure 8.
The upflow is linearly accelerated until a height of 150 Mm, and the velocity gradient becomes much steeper beyond 150 Mm.
Roughly speaking, the temperature is increasing linearly in the entire region.
The 2 MK plasma is located at 170 Mm. 
The cross sectional area expands almost proportional to the height from the solar surface, and it reaches around 25 at 150 Mm.
Because the velocity and cross sectional area are almost proportional to the height, the density is decreasing with height ($\sim s^{-2}$).
The density ratios at 100 and 170 Mm are $\sim$ 0.015 and 0.004, respectively.
It is also remarkable that the temperature at 100 and 170 Mm are $\sim$ 1.4 and 2 MK, respectively.

\subsubsection{ Weak dependence on temperature: Case III ($\nu=1.5$)}
The increase of cross sectional area between 0.4 and 1.6 MK plasma should be around $\sim 1.6$, which we already mentioned in \S 2.  
It seems that the cross sectional area expands too much in the previous case (Case II).
The above parameters are derived from the quiet sun observations.
Therefore those may be different when we discuss the cross sectional area in the dimming region. 
Next, we assumed the parameters as $T_h=10^6$ K, $\Gamma=31$, $\nu=1.5$ for the relatively gradual expansion in the cross sectional area (Case III).
The results of height dependence in the dimming region are shown in Figure 10.
The figure format is the same as Figure 8.
The characteristics in Figure 10 are very similar to that in Figure 9.
Roughly speaking, the velocity, temperature, and cross sectional area are increasing linearly, and the density is decreasing with height ($\sim s^{-2}$ ).
The difference between Figure 9 and 10 is the absolute value of the height from the solar surface.
The maximum height is $\sim 60$ Mm in Figure 10, although that is 180 Mm in Figure 9. 
The temperatures at 40 and 55 Mm are 1.4 and 2 MK, respectively.
The density ratios at 40 and 55 Mm are 0.063 and 0.03, respectively.
Note that our result is not sensitive to $T_h$ or $\Gamma$ but  sensitive to $\nu$ largely.
We have tested our method with other parameters (not shown here).

One may think that the increase of cross sectional area between 0.4 and 1.6 MK plasma is still large when we assume $\nu=1.5$.
Assuming a much smaller value in $\nu$ ($\sim 1$), the increase will become much smaller, and the height from the solar surface becomes much lower.
However, when we assume $\nu \sim 1$, there are double valued solutions in the higher temperature range.
Thus we cannot obtain the realistic solution with $\nu \sim 1$.

\subsection{Comparison between Cases I--III}
Let us discuss which is the most realistic result within our three examples.
One of the biggest differences between the three results is the height from the solar surface.
In Case I, 1.4 MK plasma (\ion{Fe}{12}) is located around 7.5 Mm from the solar surface.
On the other hand, 1.4 MK plasma is located around 100/40 Mm from the solar surface in Case II/III, respectively.
Unfortunately, we do not have any observational information of the height in this event.
However, some similar events which occurred in the limb showed that the line formation height of \ion{Fe}{12} (1.4 MK plasma) is $\lesssim$10 Mm.
\cite{bar} discussed the flare event on 2006 December 6.
The flare occurred in the same active region as our event (NOAA 10930), and their characteristics are also very similar to our event (e.g., eruptive flare, GOES class, CME associated flare).
They also found the upflow at the boundary of active region and dark region (transient coronal hole) in \ion{Fe}{12} (see Figure 16 in \cite{bar}).
The upflow was observed only in the base of a fluxtube ($\lesssim$ 10 Mm from the surface).
These findings support our results in Case I.
Further the increase of the cross sectional area also support Case I.
The observation in \cite{jin} indicated that the increases of cross sectional area between 0.4 and 1.6 MK plasma seems to be 1.6.
Therefore, we conclude that the results in Case I is the most realistic.

\section{Time-dependent Ionization}
We have discussed the height variation of temperature-dependent upflow in the dimming region by using the steady hydrodynamic equation.
To derive the relationship between the temperature and velocity from the observations, we assumed ionization equilibrium.
Generally fast flow and rapid heating may cause non-equilibrium ionization, and so far many papers discussed the time-dependent ionization with the models \citep[e.g.,][]{dup, rea, ko, ima4} and observations \citep[e.g.,][]{kat, ima3, mur}.
It is plausible that the ionization equilibrium assumption is violated in our situation, because both the temperature and velocity rapidly increase with height.
In this section we try to evaluate the ionization equilibrium assumption in our result.

In order to study the effect of transient ionization, we have calculated the time evolution of ion charge states. In our EIS observations, most of emission lines are from iron. Therefore, we concentrated on the time-dependent ionization of iron in this paper.
The method we use to calculate the time-dependent ionization is the same as that described in \cite{ima4}.
The continuity equations for iron are expressed as follows;
\begin{equation}
\frac{\partial n^{Fe}_i}{\partial t}+\nabla \cdot n^{Fe}_i {\bf v}  = 
n_e\left[n^{Fe}_{i+1} \alpha^{Fe}_{i+1}+ n^{Fe}_{i-1} S^{Fe}_{i-1}-n_i^{Fe}\left(\alpha^{Fe}_{i}+S^{Fe}_{i}\right)\right],
\end{equation}
where $n_i^{Fe}$ is the number density of the $i$th charge state of the iron, $\alpha^{Fe}_i$ represents the collisional and dielectronic recombination coefficients, and $S^{Fe}_i$ represents the collisional ionization coefficients. The ionization and recombination rates were calculated using \cite{arn1}, \cite{arn2}, and \cite{maz}. Here we assumed that all ions and electrons have the same flow speed and temperature in the same position. The ionization and recombination coefficients ($\alpha$ and $S$) strongly depend on electron temperature and weakly depend on density. The timescale for ionization and recombination is proportional to $n_e^{-1}$. 

To solve the time-dependent ionization, we need the absolute value of density.
We set the density at the base of fluxtube ($n_0$) to $10^{10}$ and $10^{11}$ cm$^{-3}$.
These values are reasonable for the density in the upper chromosphere or lower transition region \citep[e.g.,][]{ver}.
Figure 11 shows the height variation of the iron charge states  in the super radial expansion model (Case I; Figure 8).
From the top, the result of the ionization equilibrium calculation, the time-dependent ionization calculation with the lower density assumption ($n_0=10^{10}$ cm$^{-3}$), and the time-dependent ionization calculation with the higher density assumption ($n_0=10^{11}$ cm$^{-3}$) are shown.
The peak abundances of each charge state with the ionization equilibrium assumption are shown with vertical dotted lines.
From \ion{Fe}{8} to \ion{Fe}{10}, the ionization equilibrium assumption seems to be reasonable.
On the other hand, it is clearly seen that the ionization equilibrium assumption is violated from \ion{Fe}{11} to \ion{Fe}{15} in the case of lower density assumption. 
This is because there is a steep temperature and velocity gradient around 7 Mm. 
Even in the case of higher density assumption, the ionization equilibrium assumption is violated in \ion{Fe}{14} and \ion{Fe}{15}.

We also perform the same analysis in the empirical rapid expansion model (Case II; Figure 9).
Figure 12 shows the height variation of the iron charge states  in Case II.
The figure format is the same as Figure 11.
It is plausible that the ionization reaches the equilibrium stage in this mode, because the gradient of temperature and velocity in space is relatively gradual compared with the previous model.
However, even in this case, the ionization equilibrium assumption is violated in \ion{Fe}{14} and \ion{Fe}{15}.
This is because the rapid expansion causes the low density in the high temperature range, and the timescale of ionization increases.
Therefore, in either case, the ionization equilibrium assumption may be violated in the higher temperature range.

The violation of ionization equilibrium naturally causes the increasing of the line formation temperature.
For example, the peak abundances of \ion{Fe}{12} in the case of ionization equilibrium (top of Figure 11) is located around 7.5 Mm (yellow dotted line).
Thus, \ion{Fe}{12} represents $\sim$1.4 MK plasma, because the temperature at 7.5 Mm is 1.4 MK in Figure 8.
However, the peak abundances of \ion{Fe}{12} in the case of time-dependent ionization (middle of Figure 11) is located around 8 Mm.
This indicates that \ion{Fe}{12} represents $\sim$ 1.7 MK plasma, because the temperature at 8 Mm is 1.7 MK in Figure 8.
In the higher temperature range, the line formation temperatures systematically increase by considering time-dependent ionization. 
Therefore, some modification may be needed to the relationship between the velocity and temperature in Figure 5, especially in the higher temperature range.
The relationship may become close to linear after considering the time-dependent ionization.   

\section{Summary and Discussion}

We have discussed the temperature-dependent upflow in the dimming region observed by {\it Hinode}/EIS.
We have reanalyzed the dimming event on 2006 December 13, and described the previous  results for this event.
The observations show;
1) the intensity in the dimming region dramatically drops within 30 minutes from the flare onset, and the dimming region reaches the equilibrium stage after $\sim$1 hour later, 
2) the temperature-dependent upflows are observed in the dimming region during the equilibrium stage, 
3) the area of the dimming region also weakly depends on the temperature (the expansion factor is $\sim$1.6 between 0.4 and 1.6 MK plasma). 
We carried out the reconstruction of  the temperature-dependent upflow by using our new method with the assumption that both the temperature and velocity depend on the height from the solar surface.
The height variations of plasma conditions in the dimming region are demonstrated for three different cases, and we discussed which result is the most realistic. 
What we found from the reconstruction is as follows;
1) the super radial expansion of the cross sectional area is required to satisfy the mass conservation and momentum equations,
2) the temperature and velocity gradient with height become steep around 7 Mm from the solar surface, 
3) the ionization equilibrium assumption may be violated in the higher temperature range.
What we found is summarized in Figure 13.

In our analysis, we assumed that both the temperature and velocity depend on the height from the solar surface.
We already mentioned that the other scenario (the multi-strand model) can also explain the observation.
Note that it is still consistent even if we apply our result to the multi-strand model, because the higher temperature loop ($\sim 2$MK) also contains the cold component ($\sim 1$MK) at the base of the loop.
The difference between two scenarios is that the heating and acceleration are stopped before reaching to 2MK in some loops.

Let us discuss the impact of temperature gradient to the flow acceleration.
To understand intuitively, we assume $p\propto e^{-s}$ and $T = \lambda_1 s + \lambda_2$, where $\lambda_{1,2}$ is the positive parameters.
After substituting $p\propto e^{-s}$ and $T$ to Equation 3, we can derive $n\propto (\lambda_1s+\lambda_2)^{-1}e^{-s}$.
Then we found that the first term in right-hand side  in Equation 4 ($(-1/mn) dp/ds$) is proportional  to $\lambda_1s+\lambda_2$.
Roughly speaking, the second term in right-hand side in Equation 4 is negligible.
Therefore, the steep temperature gradient causes the rapid acceleration.
The steep temperature gradient causes the steep density drop.
This is the reason why the strong temperature gradient causes the strong acceleration.
Figure 14 shows each term in Equation 4 in Case I, and we can find that the trend of $(-1/mn) dp/ds$ is very similar to that of the temperature in Figure 8. 

We also perform the time-dependent ionization calculation in our models, and found that the ionization equilibrium assumption may be violated in the higher temperature range.
This naturally causes the increasing of the line formation temperature, and we need some modification to the relationship between the velocity and temperature in Figure 5.
Further, the modification causes some another modification in our reconstruction of the dimming region.
It may be possible to find the reconstruction solution which can satisfy both the hydrodynamic equation and the time-dependent ionization.
However the time-dependent ionization process heavily depends on the absolute value of density.
Unfortunately, we do not have any pairs of emission lines for density diagnostics in our observation.
Thus we cannot conclude that the modification is really needed or not at this stage.

We now discuss the energy balance in our model.
The energy equation in quasi-steady flow can be written as follows;
\begin{equation}
\frac{3}{2}vp\frac{1}{T}\frac{d T}{d s}+p\frac{d v}{d s}+vp\frac{1}{A}\frac{d A}{d s}=
\frac{1}{A}\frac{d}{d s}\left(\kappa_0AT^{\frac{5}{2}}\frac{d T}{d s}\right)-n^2\Lambda(T)+H
\end{equation}
where $\kappa_0$, $\Lambda(T)$, and $H$ are the thermal conductivity along the magnetic field line, the radiative loss function, and the heating function.
From our reconstruction of the flow in the dimming region, we can derive most of the terms in Equation 14.
The exceptions are the absolute value of density and the heating function.
Thus we can estimate the heating function with the assumption of density at the base of fluxtube.
We have calculated the heating function with $n_0=2.5\times10^{10}$ cm$^{-3}$ in Case I.
Figure 15 shows the energy balance in the dimming region.
The solid (black), dashed (blue), dotted (red), and dotted-dashed (green) lines show the height variation of the heating function, radiative cooling, thermal conduction, and left-hand side of Equation 14, respectively.
At the bottom of the fluxtube, the radiative cooling is dominated because of the low-temperature and high-density plasma.
Therefore the heating function should be large enough to maintain energy balance.
There is a steep gradient in the velocity and temperature between 6 and 7 Mm in Figure 8.
Thus the enthalpy transport and thermal conduction are dominated in this region.
Above the 7 Mm, the temperature gradient becomes gradual.
This naturally causes the rapid reduction of the thermal conduction, although the enthalpy transport reduces gradually.
Thus an increase in the heating function is required.
Our result in Case I suggests that the strong heating function may be located above 7 Mm.
Note that most of Equation 14 is dependent on the absolute value of density, and the dependence on the density is different in the each term; radiative cooling $\propto n^2$, left-hand side $\propto n^1$, thermal conduction $\propto n^0$.
Unfortunately we do not have the observational information of the density.
Thus, we do not discuss the absolute value of heating function at this stage.

Density diagnostics are crucial for the modeling of the dimming region.
Consistent modeling, taking care of the time-dependent ionization and energy equation, of the dimming region with the density diagnostics is for future work.
Another important question is the stability of the dimming flow in the category of hydro and/or magnetohydro dynamics.
So far, the stability of coronal loops in many kind of instability were tested \citep[e.g.,][]{par,hoo,zwe}.
The dimming flow seems to persist for a long period. 
The steady solutions dramatically change in some cases when the conditions are slightly modified.
Further, the critical point seems very close to the solar surface, although it is not clear whether the flow pass through a critical point or not.
It is physically important whether the flows become super sonic or not.
This may also affect the stability of the dimming region.
The stability analysis of dimming region is also future work.

\acknowledgments 
Hinode is a Japanese mission developed and launched by ISAS/JAXA, collaborating with NAOJ as a domestic partner, NASA and STFC (UK) as international partners. Scientific operation of the Hinode mission is conducted by the Hinode science team organized at ISAS/JAXA. This team mainly consists of scientists from institutes in the partner countries. Support for the post-launch operation is provided by JAXA and NAOJ (Japan), STFC (U.K.), NASA (U.S.A.), ESA, and NSC (Norway).

 This work was partially supported by the Grant-in-Aid for Young Scientists Start-up (21840062), by the Grant-in-Aid for Scientific Research B (23340045), by the JSPS Core-to-Core Program (22001),  by the JSPS fund \#R53 (''Institutional Program for (Young Researcher Overseas Visits,'' FY2009-2011) allocated to NAOJ, and by the NINS Inter-institute collarborative program for Creation of New Research Area (Head Investigator: T. Watanabe). We also acknowledge partial support by the Center for Magnetic Self Organization (NSF PHY0821899) and the hospitality of University of Wisconsin.

\begin{table*}
\begin{center}
\caption{Fitting results. }
\begin{tabular}{ccc}
\tableline\tableline
$a_2$ (K s$^2$ m$^{-2}$) &$a_1$(K s m$^{-1}$)& $a_0$ (K)   \\
\tableline
-6.00$\times$10$^{-5}$   &2.27$\times$10$^{1}$&-2.05$\times$10$^{4}$\\
\tableline
\end{tabular}
\end{center}
\end{table*}


\begin{figure}
\epsscale{0.7}
\plotone{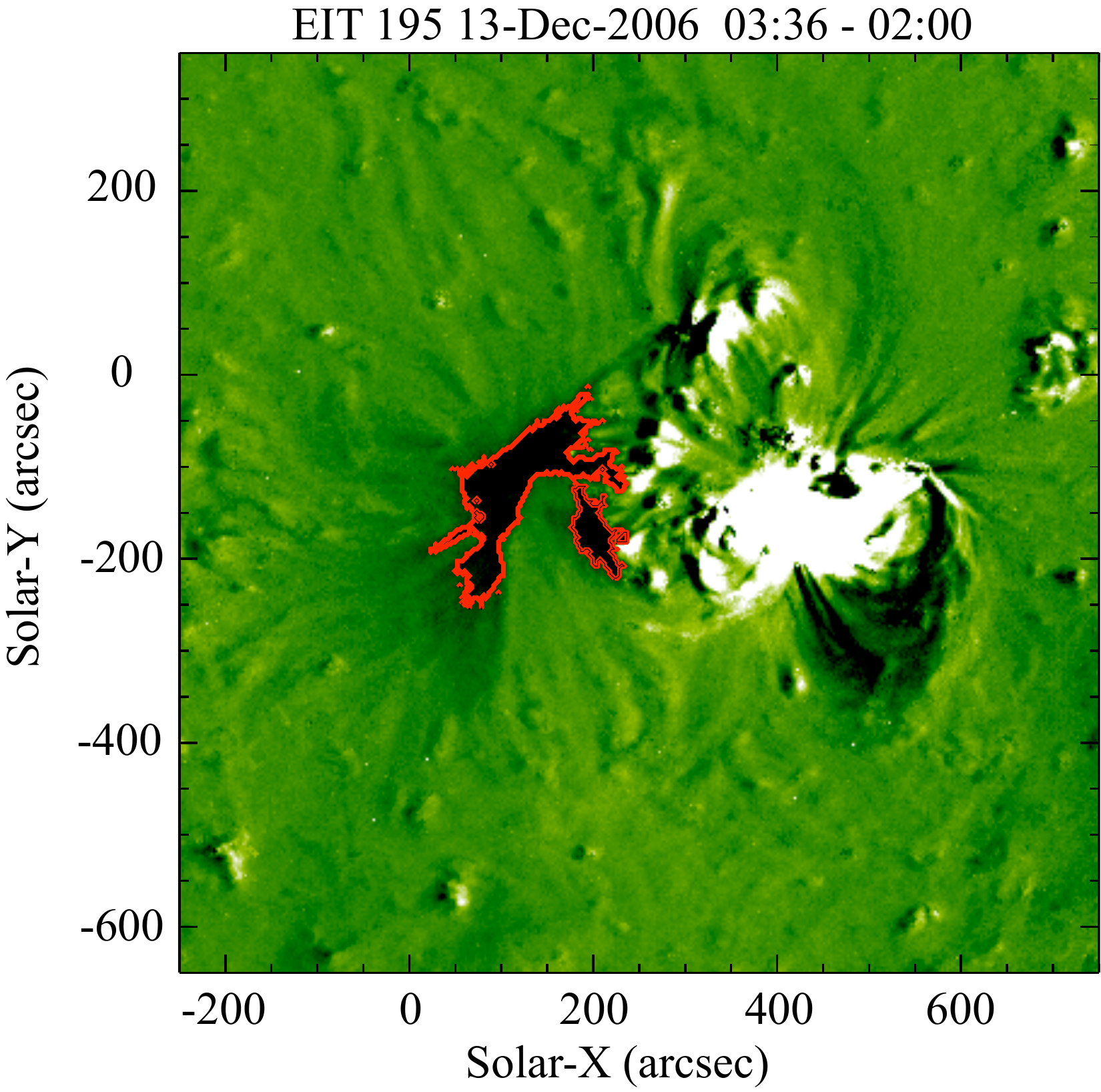}
\caption{ EIT base difference image with preflare image at December 13, 02:00 UT subtracted from the image on December 13, 03:36 UT. The red contour represents the dimming region. }
\end{figure}
\begin{figure}
\epsscale{0.7}
\plotone{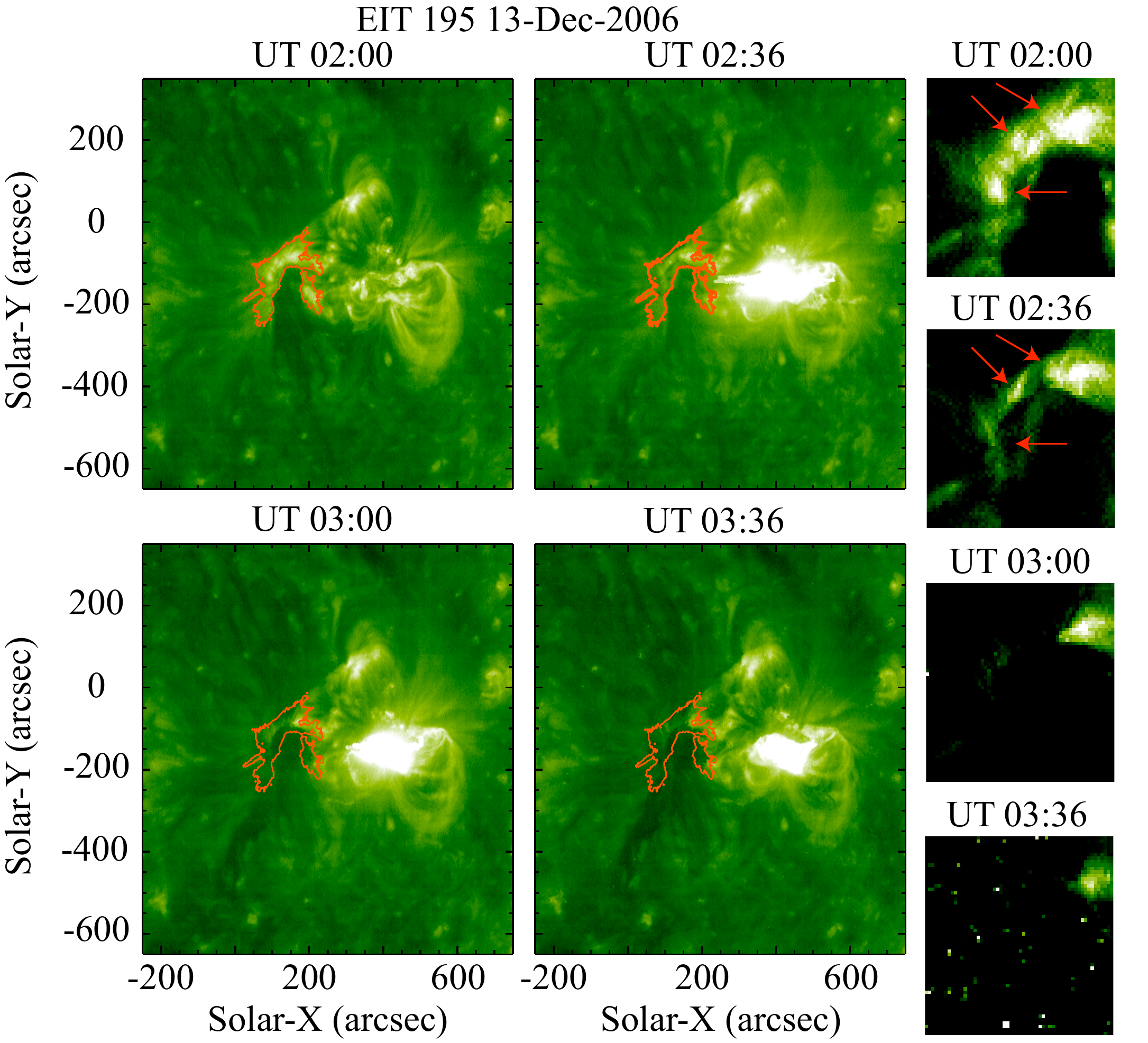}
\caption{ Time series of the dimming region. The right figures are the magnified figure of the dimming region. The red arrows show the bright region inside the dimming region. The color scale was changed from the left figures. }
\end{figure}
\begin{figure}
\epsscale{0.7}
\plotone{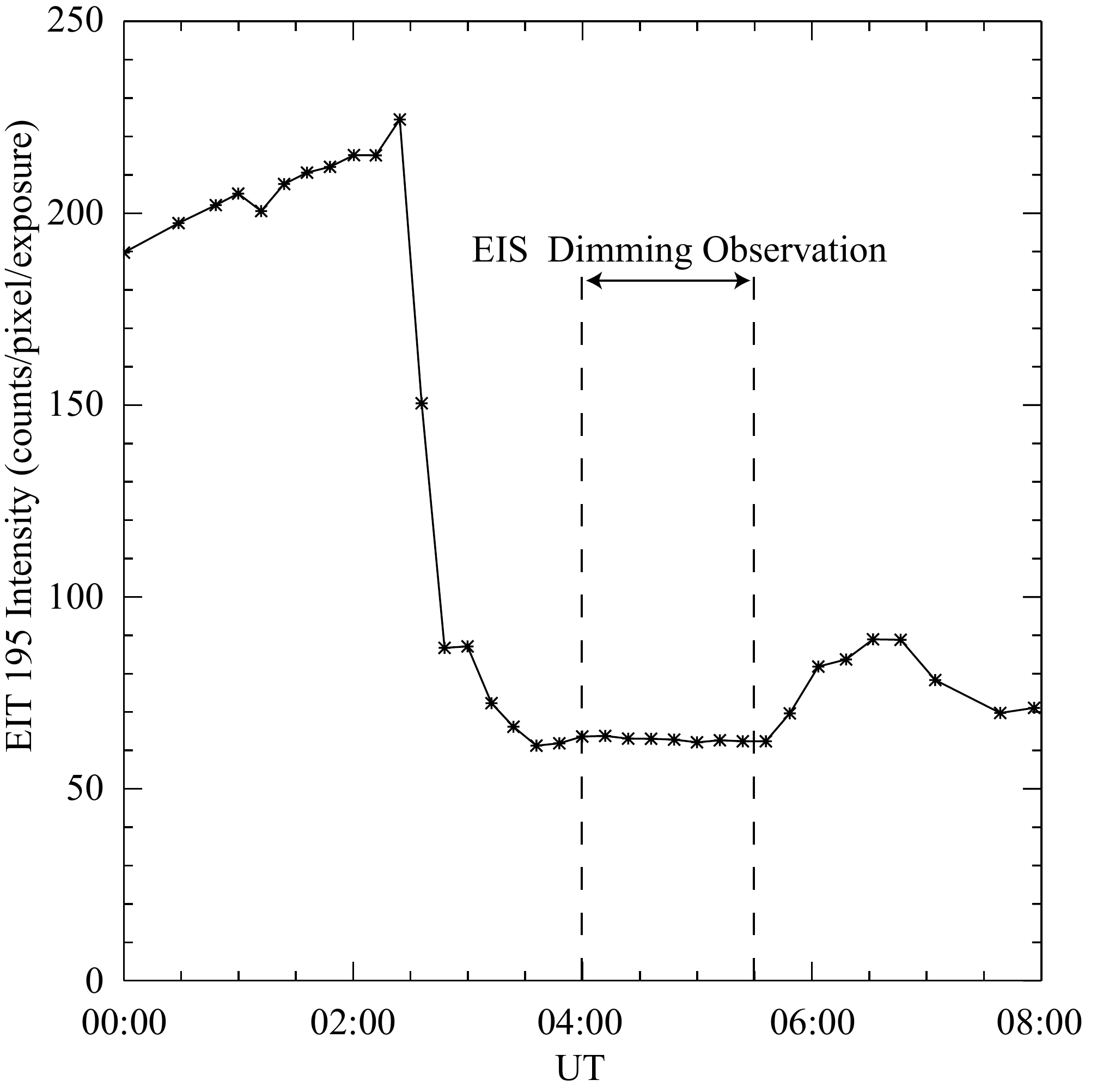}
\caption{ Time variation of the intensity in the dimming region (inside the red contour in Figure 1). The EIS observing time of the dimming region is also shown.}
\end{figure}
\begin{figure}
\epsscale{1.0}
\plotone{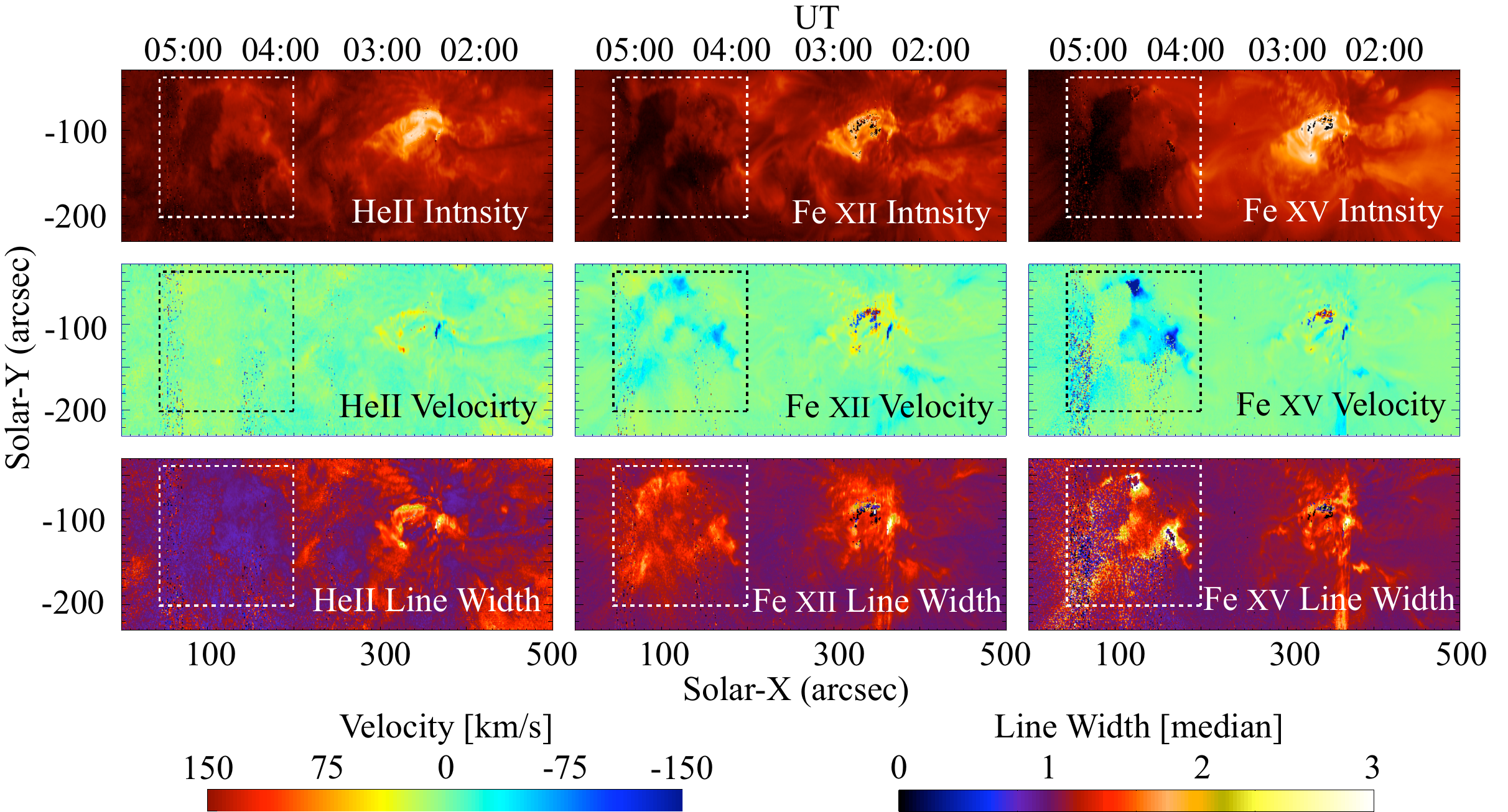}
\caption{ Intensity, velocity, and line width map of the dimming region observed by Hinode/EIS. The color scales in the velocity map range from -150 to 150 km s$^{-1}$. The color scales in the line width (Full Width at Half Maximum) map are normalized by the median of line width in whole field of view. The median line widths of \ion{He}{2}, \ion{Fe}{12}, and \ion{Fe}{15} are 0.102, 0.073, and 0.085 \AA, respectively.}
\end{figure}
\begin{figure}
\epsscale{0.7}
\plotone{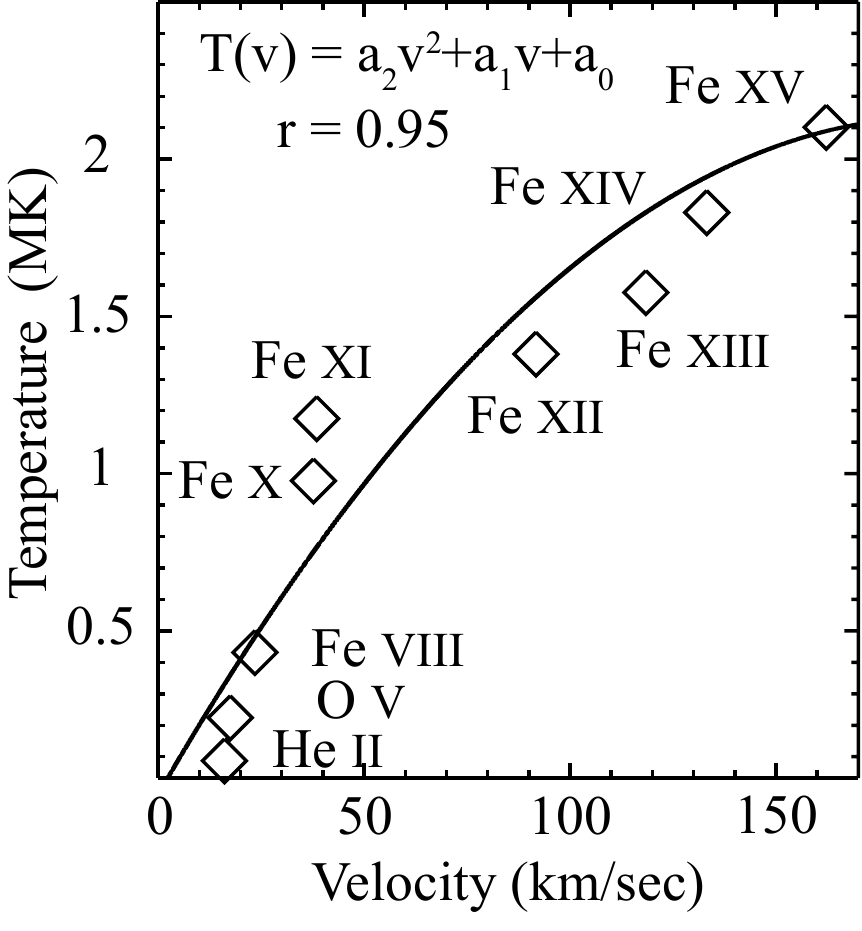}
\caption{ Relationship between temperature and velocity in the dimming region. The velocities were estimated by Doppler velocities, and the line formation temperatures were derived from ionization equilibrium assumption. }
\end{figure}
\begin{figure}
\epsscale{0.7}
\plotone{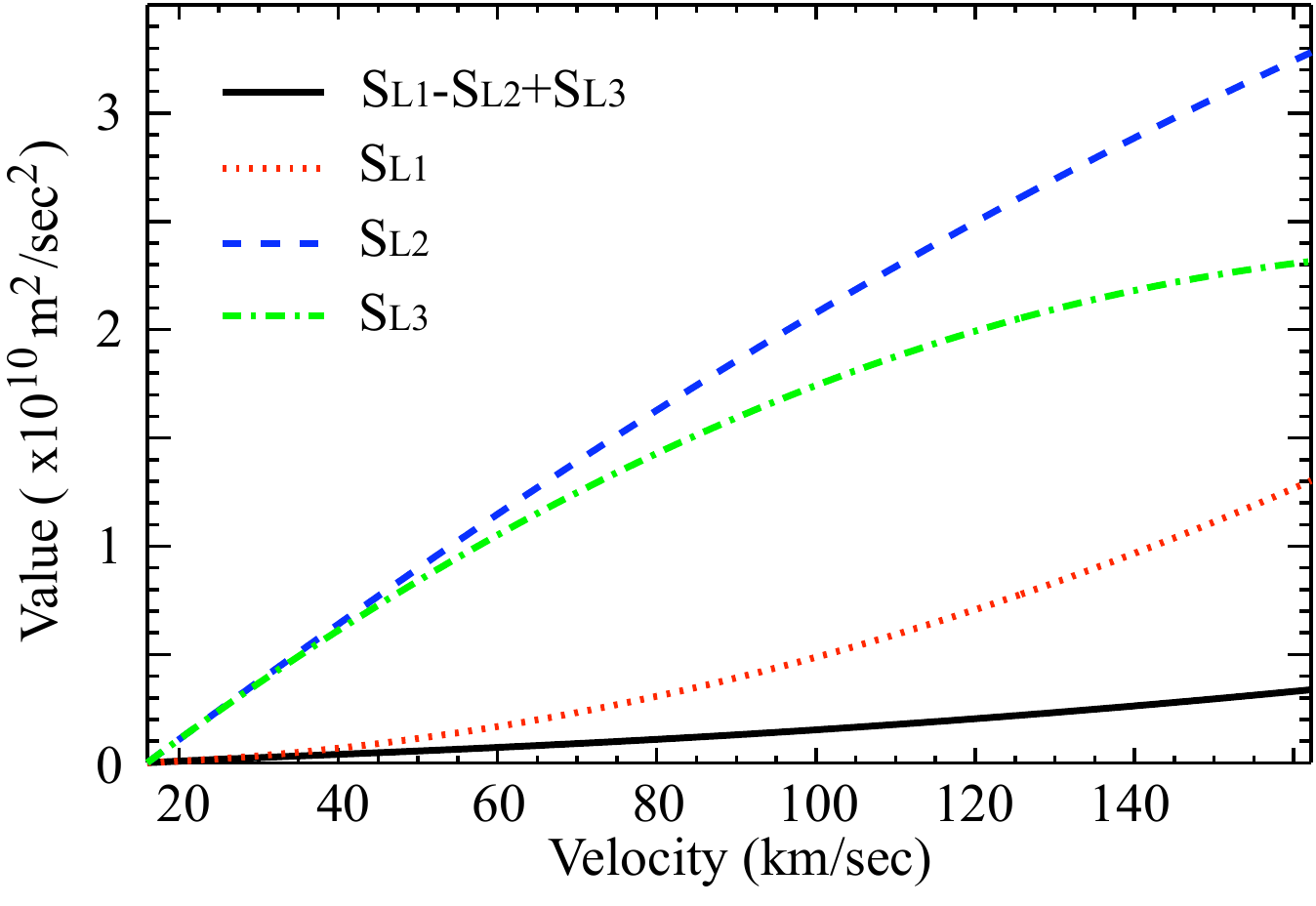}
\caption{Each terms in the left-hand side as a function of velocity in Equation 6. Solid, dotted, dashed, and dash-dotted lines show $S_{L1}-S_{L2}+S_{L3}, S_{L1}, S_{L2}$, and $S_{L3}$, respectively. }
\end{figure}
\begin{figure}
\epsscale{0.7}
\plotone{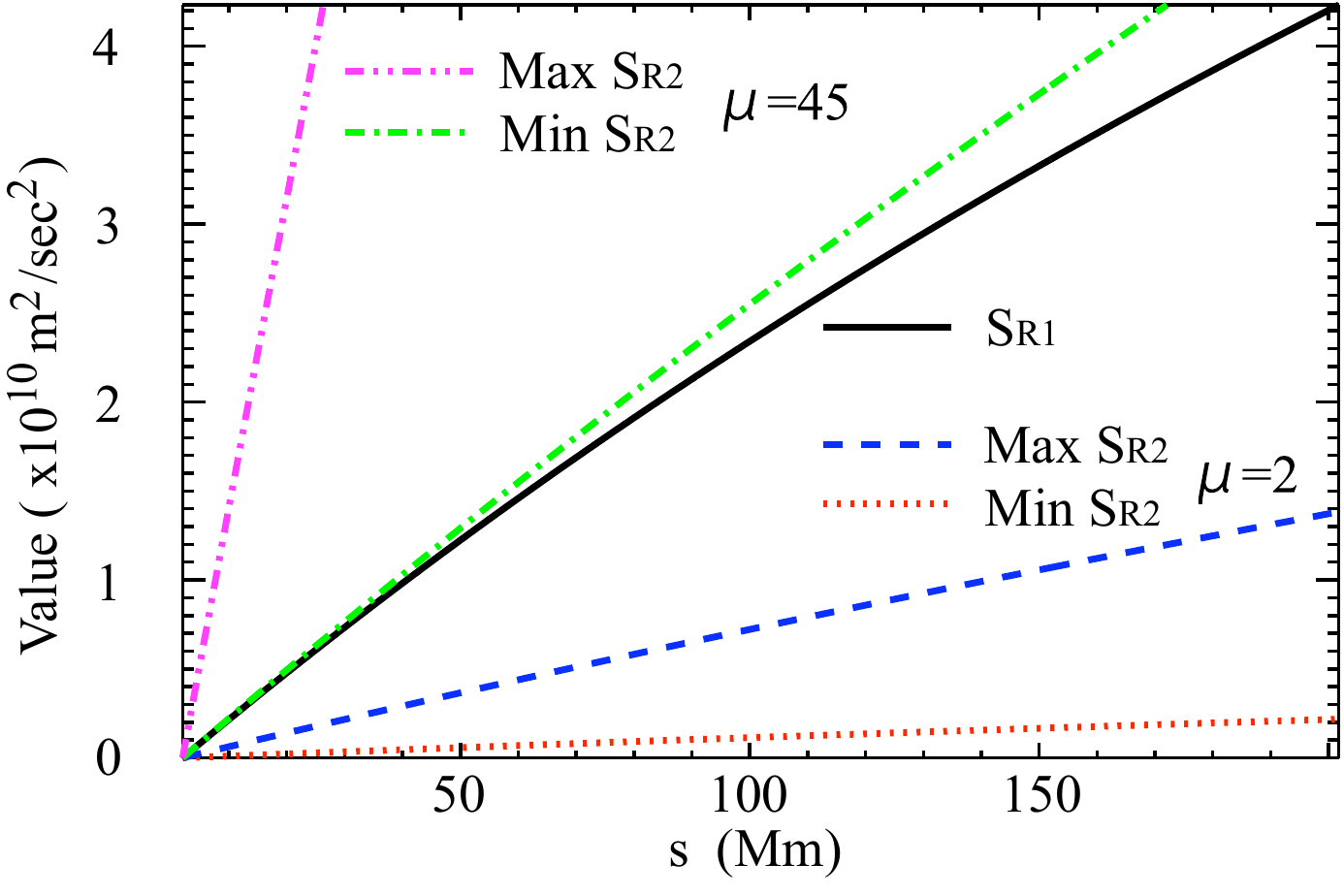}
\caption{Each terms in the right-hand side as a function of height from the solar surface in Equation 6. }
\end{figure}
\begin{figure}
\epsscale{0.5}
\plotone{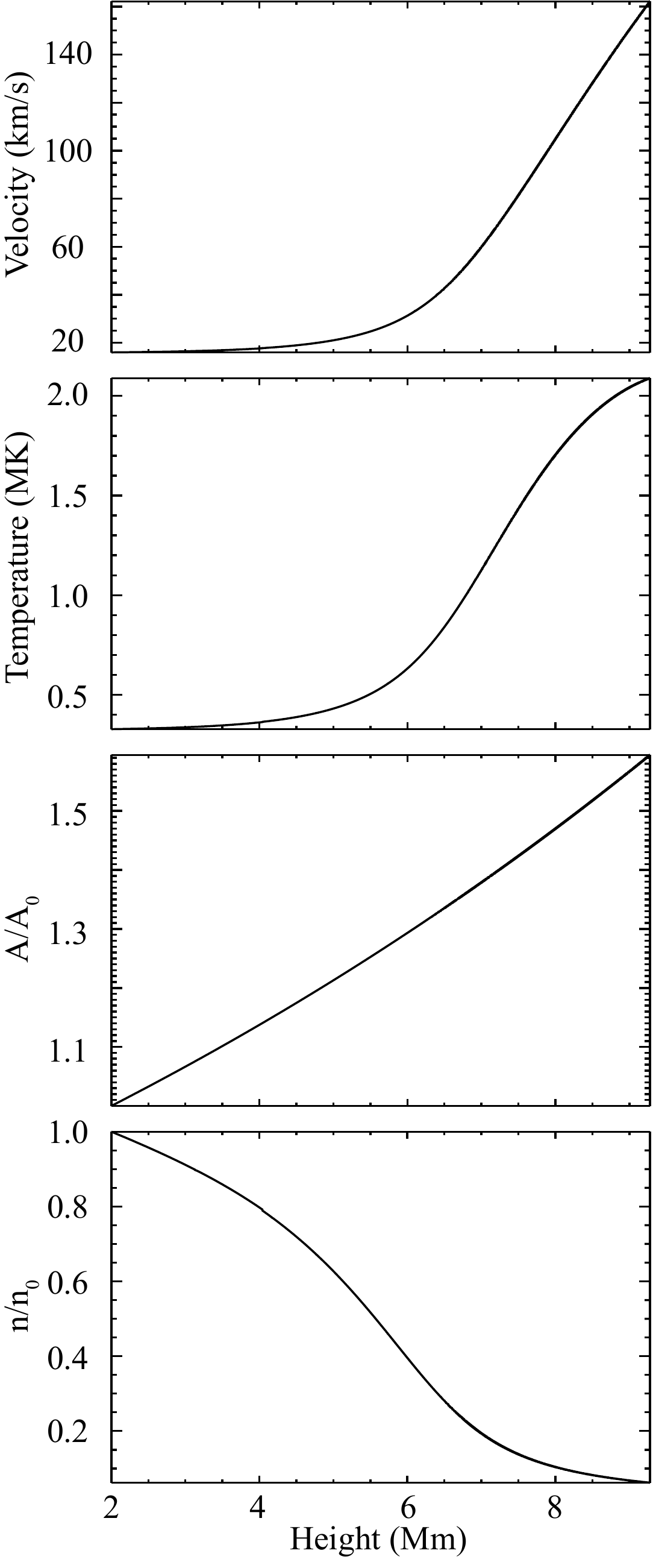}
\caption{ Height variation of plasma conditions in the case of $\mu=45$ (Case I: super radial expansion). From the top to bottom, velocity, temperature, increase of cross section, and density drop are shown. The cross sectional area and density are normalized by that at $s=s_0$.}
\end{figure}
\begin{figure}
\epsscale{0.5}
\plotone{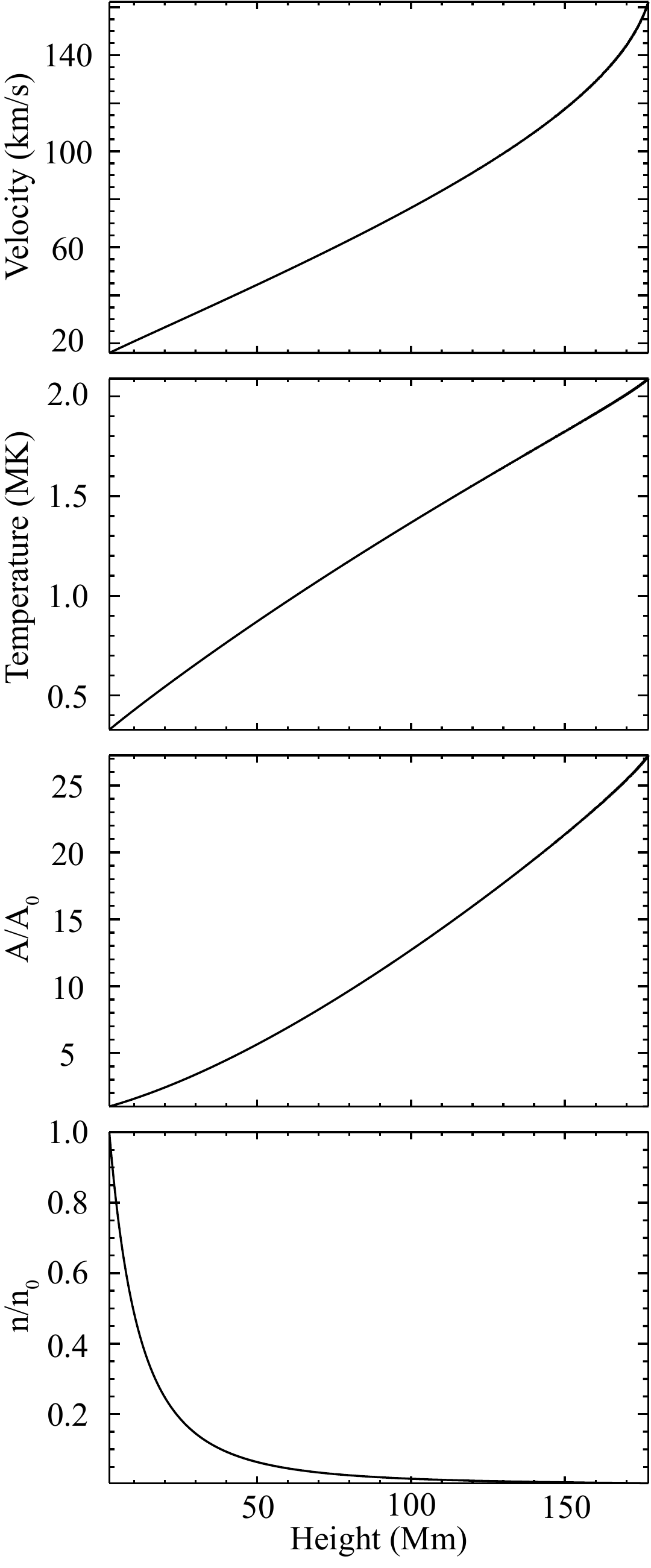}
\caption{Height variation of plasma condition in the case of $\nu=3.6$ (Case II: empirical strong expansion model). The figure format is the same as Figure 8.}
\end{figure}
\begin{figure}
\epsscale{0.5}
\plotone{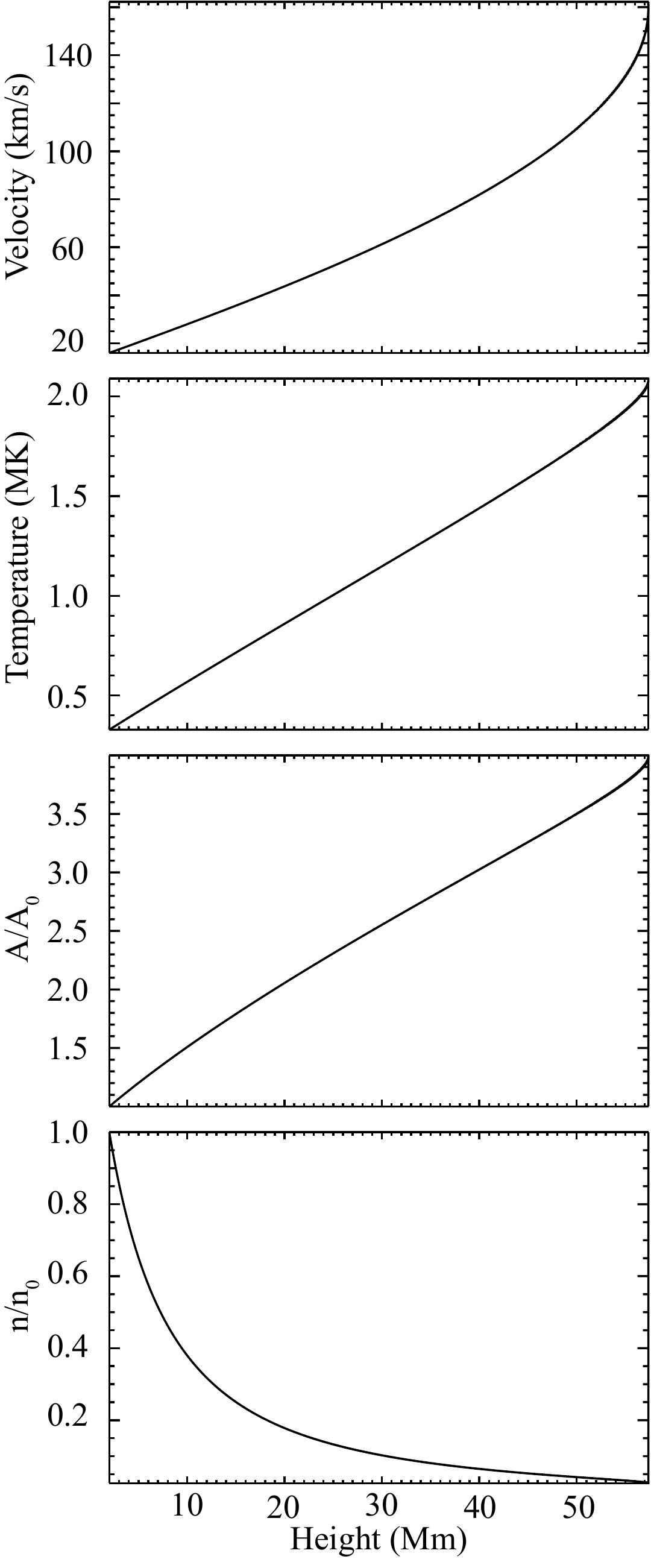}
\caption{ Height variation of plasma condition in the case of $\nu=1.5$ (Case III: empirical weak expansion model). The figure format is the same as Figure 8.}
\end{figure}
\begin{figure}
\epsscale{0.5}
\plotone{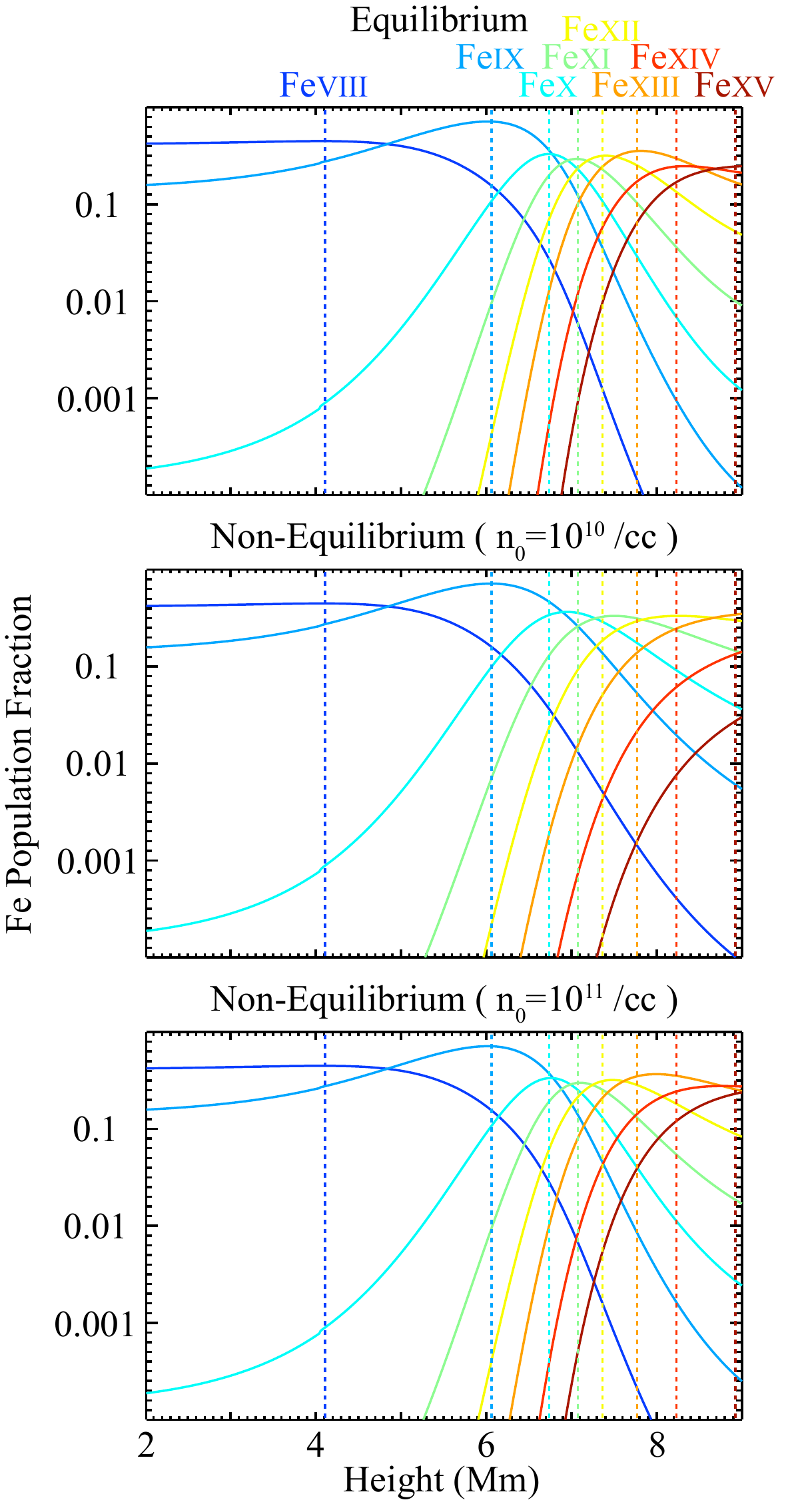}
\caption{ Time-dependent ionization in the case of $\mu=45$ (Case I: super radial expansion). From the top to bottom, the results with ionization equilibrium, non-equilibrium ($n_0=10^{10}$ cm$^{-3}$), non-equilibrium ($n_0=10^{11}$ cm$^{-3}$) are shown, respectively. The vertical dotted lines show the heights of peak abundance in each elements in the ionization equilibrium results (Top).}  
\end{figure}
\begin{figure}
\epsscale{0.5}
\plotone{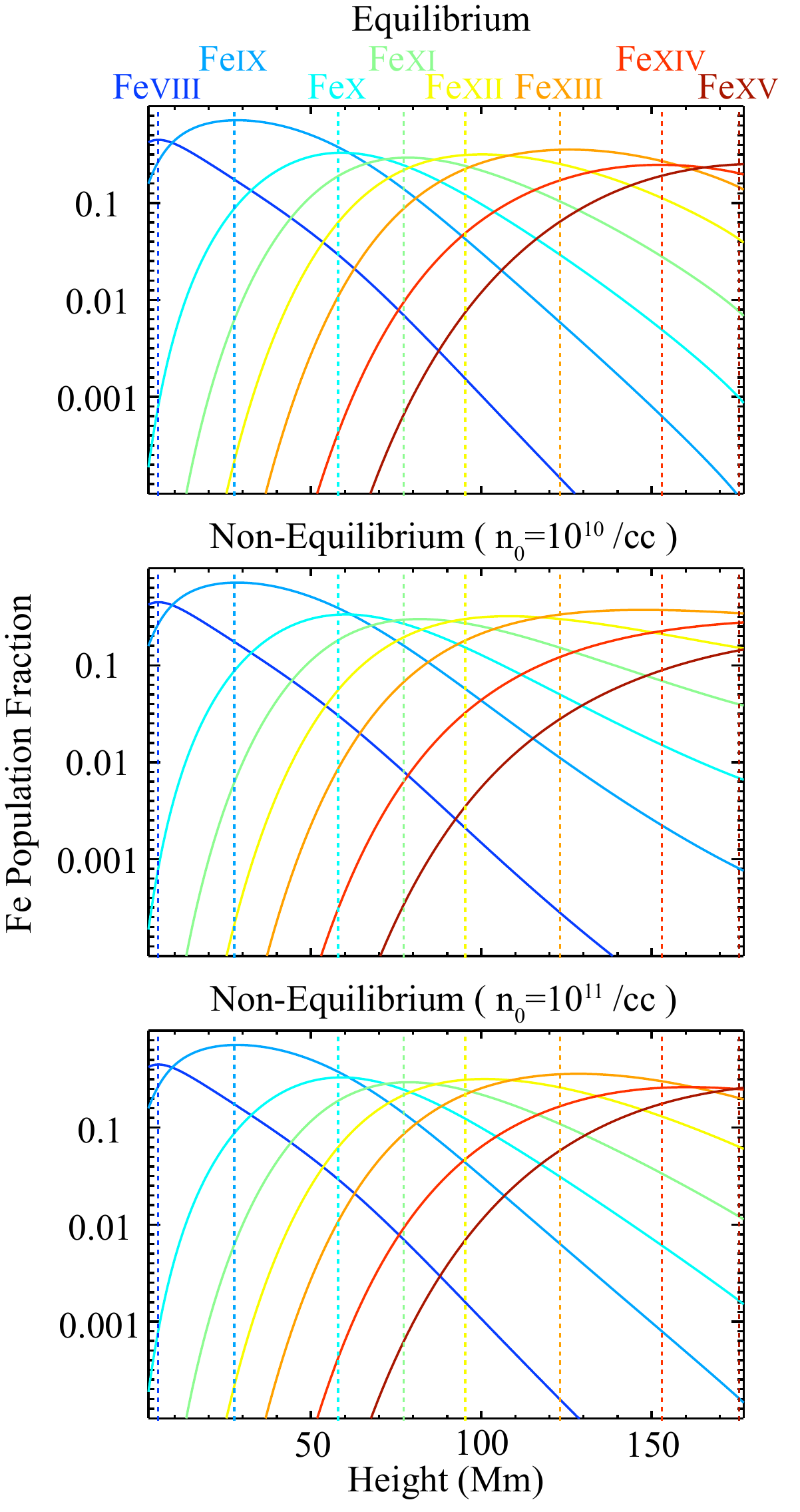}
\caption{ Time-dependent ionization in the case of $\nu=3.6$ (Case II: empirical strong expansion model). The figure format is the same as Figure 11.}
\end{figure}
\begin{figure}
\epsscale{0.8}
\plotone{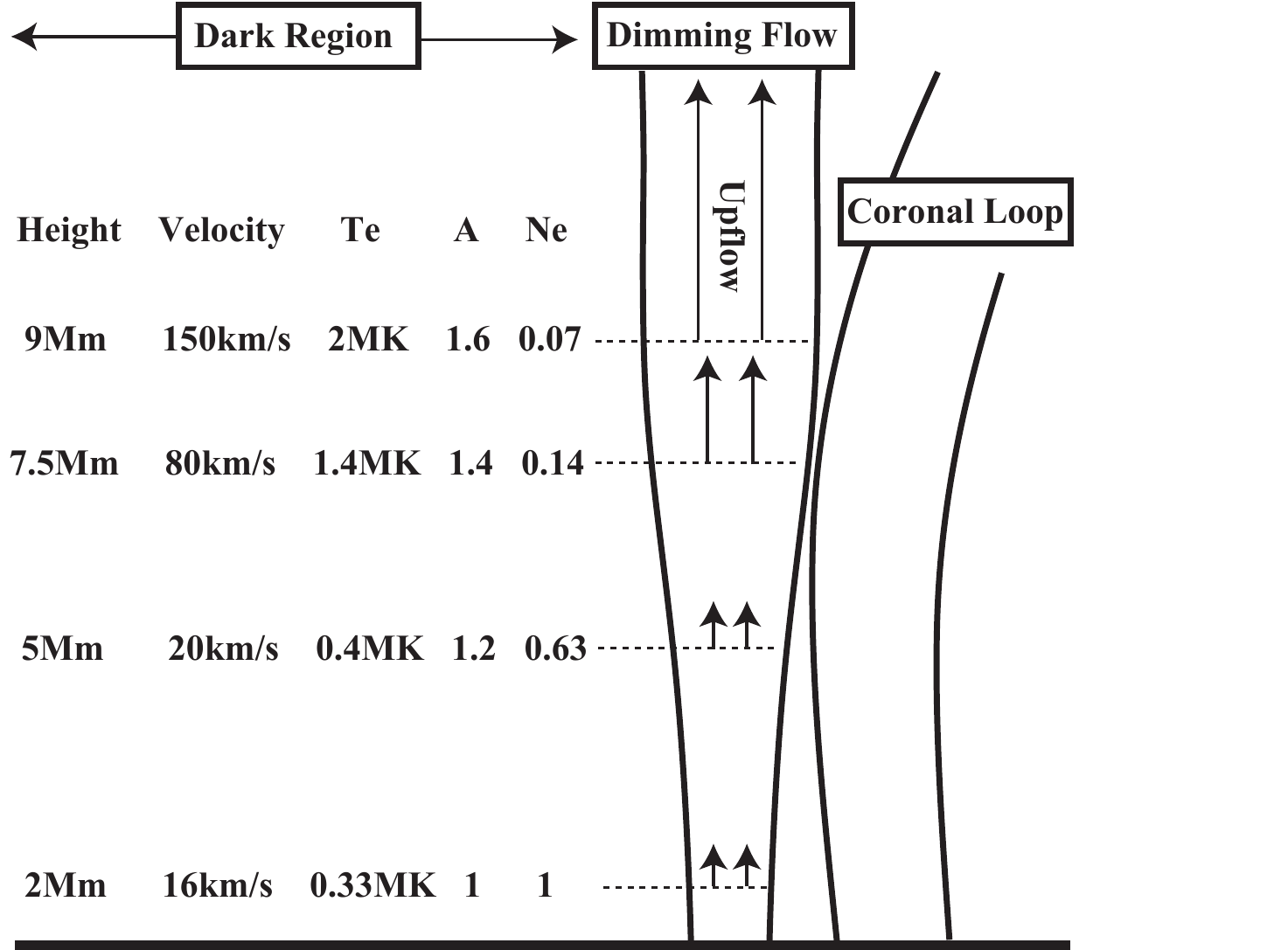}
\caption{ Schematic illustration of our observation and reconstruction in and around the dimming region. $T_e$ shows the electron temperature in the dimming region. $A$ and $N_e$ show the cross sectional area and electron density which are normalized at the base (2 Mm). }
\end{figure}
\begin{figure}
\epsscale{0.8}
\plotone{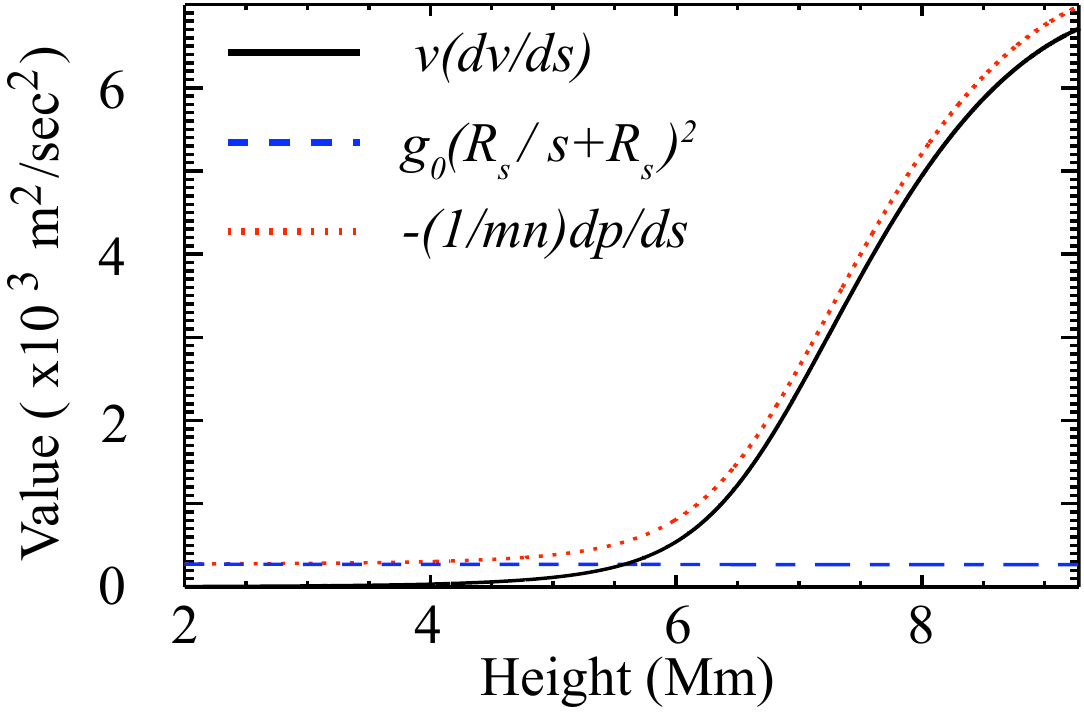}
\caption{ Each terms in Equation 4 (Case I: super radial expansion).}
\end{figure}
\begin{figure}
\epsscale{0.8}
\plotone{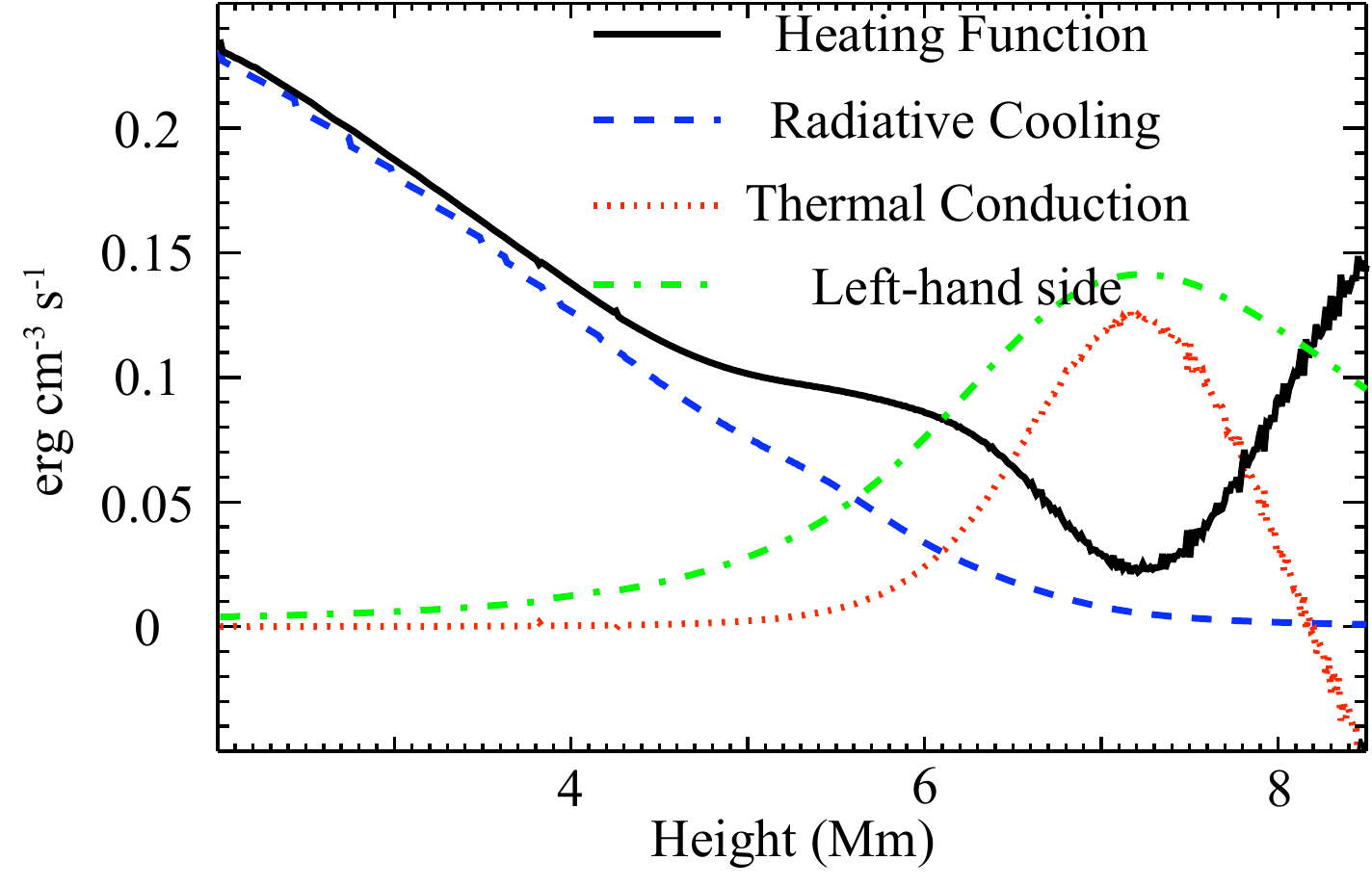}
\caption{ Energy balance in the dimming region (Case I: super radial expansion).}
\end{figure}

\end{document}